\documentclass[%
superscriptaddress,
preprint,
 amsmath,amssymb,
 aps,
]{revtex4-1}
\usepackage{float}
\usepackage{graphicx}
\usepackage{indentfirst}
\usepackage{geometry}
\def\comment#1{}
\usepackage{titlesec}
\begin{document}

\title{Gravitational wave emission under general parametrized metric from extreme mass ratio inspirals}
\author{Shuo Xin}
\affiliation{%
Shanghai Astronomical Observatory, Shanghai, 200030, China}
\affiliation{%
School of Physics Sciences and Engineering, Tongji University, Shanghai 200092, China}
\author{Wen-Biao Han}
\email{wbhan@shao.ac.cn}
\author{Shu-Cheng Yang}
\affiliation{%
Shanghai Astronomical Observatory, Shanghai, 200030, China}
\affiliation{%
School of Astronomy and Space Science, University of Chinese Academy of Sciences, Beijing, 100049, China}

\date{\today}

\begin{abstract}
	Future space-borne interferometers will be able to detect gravitational waves at $10^{-3}$ to $10^{-1}$ Hz. At this band extreme-mass-ratio inspirals (EMRIs) can be promising gravitational wave sources. In this paper, we investigate possibility of testing Kerr hypothesis against a parametrized non-Kerr metric by matching EMRI signals. However, EMRIs from either equatorial orbits or inclined orbits suffer from the ``confusion problem". Our results show that, within the time scale before radiation flux plays an important role, small and moderate deviations from the Kerr spacetime($|\delta_i|<1$) can be discerned only when spin parameter is high. In most cases, the EMRI waveforms related with a non-Kerr metric can be mimicked by the waveform templates produced with a Kerr black hole.   
\end{abstract}

\maketitle
\section{Introduction}

LIGO's detection of Black Hole (BH) merger event GW150914 has opened the era of Gravitational wave astronomy\cite{ligo}. Observation of  GW170817 \cite{170817}and its electromagnetic counterpart led us to multi-messenger astronomy\cite{multi}. Electromagnetic observation has deepened our understanding of the universe since born of astronomy and the new messenger, gravitational wave, might bring more discoveries. With ground-based detectors\cite{a_ligo}\cite{virgo}, space interferometers\cite{lisa_org} and pulsar timing arrays\cite{PTA}, we may observe merger events, EMRI, primordial GW and etc. These observations that are unaccessible to us previously could potentially deepen our understanding for the universe and fundamental physics.

Current ground-based gravitational wave detectors are able to detect gravitational waves (GWs) at relatively high frequency band. Laser Interferometer Space Antenna (LISA \cite{lisa}) , Taiji \cite{hu2017the} and TianQin \cite{luo2016tianqin}, planned to launch in 2030s, will extend the observation band down to milli-Hertz . LISA pathfinder has demonstrated desired accuracy by its noise spectrum\cite{LPF} and future LISA task can possibly release enormous scientific yields. By analyzing GW signals at LISA band, we can study the merger history of BHs\cite{lisa_mergerhistory}, probe stellar dynamics\cite{sameOmg} and test gravity theories\cite{test_GR}.

Extreme-Mass-Ratio Inspiral (EMRI), e.g. a stellar mass compact object (1-10 $M_{\odot}$) orbiting around a Supermassive Black Hole (SMBH), is a promising source of GW signal at LISA band. 
Although a relatively accurate method to generate waveforms, the Teukolsky-based method \cite{TB,review_waveform,han2010gravitational,han2011prd,han2014gravitational,han2016cqg}, is computationally expansive, some further approximations, i.e. Numerical Kludge\cite{kludge}, Analytic Kludge\cite{AK} and etc, made the calculation feasible. The analysis of EMRI signal is also sophisticated. Due to low signal-to-noise ratio (SNR), matched filtering has to be utilized in the analysis, and both generating waveform templates and matching signal require enormous computation power. Markov-Chain Monte Carlo (MCMC) and Machine Learning might shed some light on it \cite{MCMC} \cite{machine_learning}.

One scientific goal of LISA task is testing General Relativity (GR) and Kerr metric in strong field regime, e.g. around BHs. There are already several previous works demonstrating possible constraints LISA can set for alternative metric or theory of gravity. \cite{test_scalar-tensor} shows the bound of coupling constant $\omega$ in scalar-tensor theory that LISA can set. \cite{test_bumpyBH} gives the estimation of the parameter limits for ``Bumpy BH metric'' constrained by EMRI detection. Most works are using the error estimated by Fisher Matrix calculation as the constraint on the parameters. 

However, as \cite{07conf} mentioned, due to large parameter space EMRI has, ``confusion problem" can prevent us from parameter estimation and, therefore, testing alternative theory or metric. Namely, two EMRI waveforms with different parameters can be almost identical (that's to say, their overlap, as defined in \ref{p_wave}, is over 0.97). This problem is demonstrated in \cite{majorPRD} for a special case, i.e. consider an approximate metric describing BH+torus and Kerr metric, confusion problem exists for EMRI emission from equatorial orbits. We try to extend the analysis to continuously parametrized metric and inclined eccentric orbits.

In order to test No-Hair Theorem, i.e. astrophysical BHs are described only by mass and spin, many methods of model-independent parametrization for BH metric have been proposed. The JP metric proposed by Tim Johannsen and Dimitrios Psaltis\cite{johannsen} and the Johannsen metric proposed by Tim Johannsen \cite{johannsen_final}, expanding the metric component in power series of $\frac{M}{r}$, are widely used in testing Kerr hypothesis. However, as mentioned in \cite{johannsen_diff}, Johannsen metric has some convergence deficiencies in strong field regime. Such problem can be solved by the parametrization proposed in \cite {KRZ}, expanding the metric functions in power series of $\cos \theta$. Here we apply the lowest order KRZ metric and adopt the choice of deformation parameters in \cite{cosimoKRZ}, another work on testing Kerr metric.

The rest of this paper is organized as follows. In Sec. \ref{p_krz}, general parametrization for BH spacetime is discussed. Sec. \ref{p_wave} introduced the ``Kludge" waveform generation method and matched filtering procedure that we applied. Then we present our results about the ``confusion" problem for equatorial orbits and inclined orbits in Sec. \ref{p_conf}. Finally, we summarize and discuss the results in Sec. \ref{p_fin}. 

\section{General Parametrization of Metric around Black Hole}
\label{p_krz}

In order to test Kerr hypothesis or General Relativity in a model-independent manner, one usually turns to general parametrization of metric describing astrophysical BHs. Instead of using a metric derived from a specific theory, a general metric could enable model-independent test of Kerr hypothesis. One reasonable choice is to expand the metric functions in power series of $\frac 1 {r^2+a^2cos^2\theta}$, as adopted by Johannsen and Psaltis. \cite{johannsen}. Under Boyer-Lindquist coordinates, the metric, which we refer to as JP metric, reads:
\begin{equation}
\begin{aligned}
	ds^2=&-[1+h(r,\theta)](1-\frac{2Mr}{\Sigma}) dt^2 - \frac{4aMrsin^2\theta}{\Sigma} [1+h(r,\theta)] dtd\phi + \frac{\Sigma[1+ h(r,\theta) ] }{ \Delta + a^2 \sin^2 \theta h(r,\theta )} dr^2  \\
	&+\Sigma d\theta^2 +[ \sin^2\theta (r^2+a^2 + \frac{2a^2Mr\sin^2\theta }{\Sigma}) + h(r,\theta ) \frac{a^2 (\Sigma +2Mr)\sin^4\theta  }{\Sigma} ] d\phi^2
\end{aligned}
\end{equation}
where
\begin{equation}
	\Sigma=r^2 +a^2\cos^2\theta,\,\,\,\Delta= r^2 - 2Mr +a^2 ,\,\,\,  h(r,\theta ) = \sum_{k=0}^{\infty} (\epsilon_{2k} + \epsilon_{2k+1} \frac{Mr}{\Sigma}) (\frac{M^2}{\Sigma})^k
\end{equation}

When testing Kerr metric, one usually hopes that the alternative metric still preserves the symmetries of Kerr metric, which are related to three constants of motion. A general form of metric that has three constants of motion is proposed by Johannsen\cite{johannsen_final}. The line element of this parametrization in Boyer-Lindquist coordinates, which we refer to as Johannsen metric, is:
\begin{equation}
\begin{aligned}
	ds^2 =& -\frac{\tilde{\Sigma} [\Delta - a^2 A_2(r)^2 \sin^2\theta ] }{ [ (r^2+a^2)A_1(r) - a^2 A_2(r) \sin^2\theta ]^2 } dt^2 -  \frac{a [(r^2 + a^2 )A_1(r)A_2(r) - \Delta ] \tilde{\Sigma} \sin^2\theta }{ [(r^2+a^2 ) A_1(r) -a^2 A_2(r) \sin^2 \theta ]^2 } dtd\phi\\
	& + \frac{\tilde{\Sigma} \sin^2\theta [(r^2+a^2)^2 A_1(r)^2 - a^2\Delta \sin^2\theta  ]}{ [(r^2+a^2)A_1(r) - a^2 A_2(r) \sin^2 \theta ]^2 } d\phi^2 +\frac{\tilde{\Sigma}}{\Delta A_5(r)} dr^2  + \tilde{\Sigma} d\theta^2
\end{aligned}
\end{equation}

where $\Delta$ is defined in the same way as JP metric, and the functions $A_i(r),\,i=1,2,5$ and $\tilde{\Sigma}$ are expanded in power series of $\frac{M}{r}$

\begin{equation}
\begin{aligned}
	A_i(r) =& 1+ \sum_{n=2}^{\infty} \alpha_{in} (\frac M r)^n\\
	\tilde{\Sigma} =& \Sigma +f(r)\\
	f(r)=&\sum_{n=3}^{\infty} \epsilon_n \frac{M^n}{r^{n-2}}
\end{aligned}
\end{equation}

Johannsen metric and JP metric have been adopted by several works on testing Kerr metric, which utilize Ironline\cite{t1_Iron} \cite{t4_Iron}, X-ray polarization\cite{t2_XrayPol}, Black Hole shadows\cite{t3_BHShadow} and etc. However, as mentioned in \cite{johannsen_diff} and \cite {KRZ}, Johannsen metric has several deficiencies. One major problem is expanding the function in power series of $1/r$, so that all element in the series are almost equally important near horizon, which puts a burden when testing Kerr hypothesis in strong field regime. As we will discuss in Sec. \ref{p_conf}, we have to study the dynamics as close to the horizon as possible to mitigate the ``confusion''. This convergence problem can be solved by expanding the metric functions in power series of $\cos \theta$, as adopted in \cite{KRZ}. The line element around an axisymmetric black hole proposed by Konoplya, Rezzolla and Zhidenko, which we refer to as KRZ metric, is \cite{KRZ}:
\begin{equation}
\begin{aligned}
	ds^2=-\frac{N^2(r,\theta)-W^2(r,\theta)\sin^2\theta}{K^2(r,\theta)} dt^2-2W(r,\theta)r\sin^2\theta dtd\phi \\+ K^2(r,\theta) r^2\sin^2\theta d\phi^2  +\Sigma(r,\theta)(\frac{B^2(r,\theta)}{r^2 N^2(r,\theta)} dr^2+d\theta^2)
\end{aligned}
\end{equation}
Where $\Sigma$ is defined the same as JP metric and the functions $K(r,\theta),\, N(r,\theta),\, W(r,\theta),\, B(r,\theta),\ $ can be expanded in power series of $\cos\theta$. Here we use the same deformation parameter as \cite{cosimoKRZ}, namely $\delta_i, \, i=1,2,3,4,5,6,7,8$ related to the metric functions by

\begin{equation}
\begin{aligned}
N^2&=(1-r_0/r)[ 1-\epsilon_0r_0/r +(k_{00}-\epsilon_0)r_0^2/r^2 +\delta_1 r_0^3/r^3 ] \\
&+ \{ a_{20} r_0^3/r^3 +a_{21} r_0^4/r^4 + k_{21}r_0^3/r^3[ 1+\frac{k_{22}(1-r_0/r) }{1+k_{23}(1-r_0/r)} ]^{-1}   \} \cos^2\theta   \\
B&=1+\delta_4r_0^2 /r^2 +\delta_5r_0^2 \cos^2\theta /r^2\\
W&=[w_{00}r_0^2 /r^2 +\delta_2 r_0^3/r^3 +\delta_3 r_0^3/r^3 \cos^2\theta ]/\Sigma\\
K^2&= 1+aW/r+\{k_{00}r_0^2/r^2 +k_{21}r_0^3/r^3 [ 1+\frac{k_{22}(1-r_0/r) }{1+k_{23}(1-r_0/r)} ]^{-1} \cos^2\theta \}/\Sigma\\
r_0&=1+\sqrt{1-a^2},\,\,\, a_{20}=2a^2/r_0^3,\,\,\, a_{21}=-a^4/r_0^4 +\delta_6 ,\,\,\, \epsilon_0=(2-r_0)/r_0,\,\,\, k_{00}=a^2/r_0^2,\\
 k_{21}&=a^4/r_0^4 -2a^2/r_0^3-\delta_6,\,\,\, w_{00}=2a/r_0^2,\,\,\, k_{22}=-a^2/r_0^2 +\delta_7,\,\,\,  k_{23} = a^2/r_0^2 +\delta_8
\end{aligned}
\end{equation}
\begin{figure}[!ht]
	\centering
	\includegraphics[width=10cm]{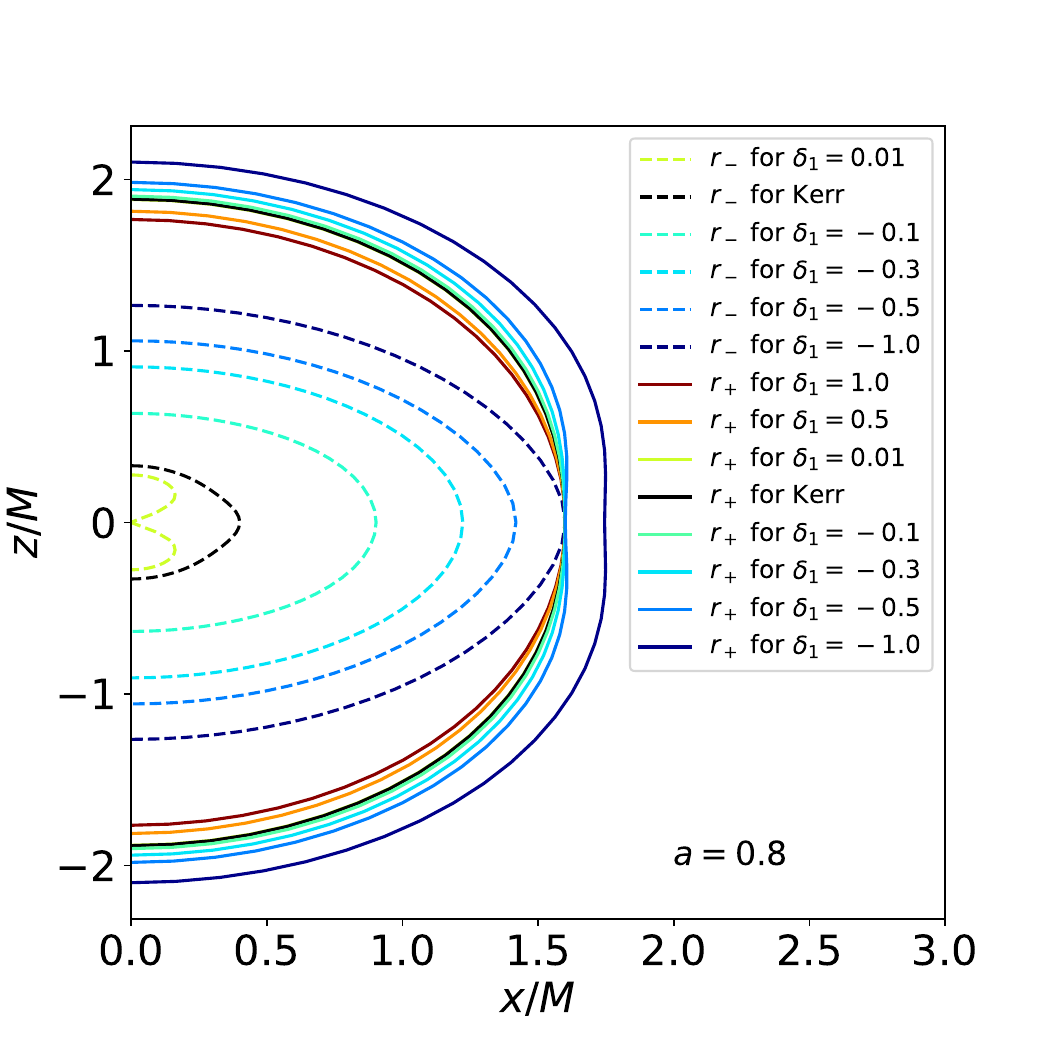}
	
	\caption{Influence of $\delta_1$ on the shape of horizon. The dashed lines of different color indicate the inner horizon $r_-$ and the solid lines indicate the outer horizon $r_+$. }
	\label{hori}
\end{figure}	
Note that here the coordinate $r$ and BH spin $a$ are redefined by $r/M$ and $a/M$ for brevity in the expression. This is a lowest order metric expression as shown in the Appendix of \cite{KRZ}, where we replace $a_{01},\,w_{01},\, w_{21},\, b_{01},\, b_{21}$ by $\delta_1,\, \delta_2,\, \delta_3,\, \delta_4,\, \delta_5$, respectively. $\delta_1$ is related to deformation of $g_{tt}$, $\delta_2$ and $\delta_3$ are related to the rotational deformation, $\delta_4$ and $\delta_5$ are related to deformation of $g_{rr}$  KRZ parametrization only preserves stationarity and axisymmetry. When each $\delta_i$ is set to 0, the metric recovers Kerr metric. In this paper we mainly consider influence of $\delta_1, \, \delta_2$. To get a sense for the influence of deformation parameters, in Fig. \ref{hori} we plot the horizon under different value of $\delta_1$ with spin parameter $a=0.8$.


\section{Kludge Waveform and Signal Analysis}
\label{p_wave}

In this section we review the Kludge waveform generation method and signal analysis approach. 

We use the method established in \cite{kludge}, i.e. Kludge waveform, to calculate EMRI signals. The procedure is: regarding the stellar mass object as a point particle, first calculate the trajectory of the particle in a given metric by integrating geodesic equations; then use quadrupole formula to get the gravitational wave from test particle geodesics.

In our instance, to calculate the geodesics, we use:
\begin{equation}
	\dot{u^\mu}=-\Gamma^\mu_{\rho\sigma}u^\rho u^\sigma
\end{equation}
\begin{equation}
	\dot{x^\mu}=u^\mu 
\end{equation}
where $x^\mu$ is the Boyer-Lindquist coordinate of the particle, $u^\mu$ is the 4-velocity and $\Gamma^\mu_{\rho\sigma}$ is Christoffel connection. We didn't use conservation of particle mass, energy and angular momentum to reduce equations but to monitor numerical error. Namely at each step of integration, we check the conservation quantities, namely the modulus of 4-velocity, energy $E$ and $z$ component of angular momentum $L_z$ defined by:
\begin{equation}
	-1 = g_{\mu\nu} u^\mu u^\nu
\end{equation}
\begin{equation}
	E = -u_t = - g_{tt} u^t -g_{t\phi} u^\phi
\end{equation}
\begin{equation}
	L_z = u_\phi = g_{t\phi } u^t + g_{\phi\phi} u^\phi
\end{equation}
 and in Kerr cases, we also check Carter constant
 \begin{equation}
 	Q = (g_{\theta\theta} u^\theta)^2 + \cos ^2 \theta (a^2 (\eta^2-E^2) + (\frac{L_z}{\sin \theta})^2 )
 \end{equation}
During the calculation, we keep the relative drift of conserved quantities within $10^{-7}$.

For stable bounded geodesics, three parameters, i.e. eccentricity $e$, semi-latus $p$ and inclination angle $\iota$, can be used to characterize an orbit. They are defined by:
\begin{equation}
\begin{aligned}
e=\frac{r_a-r_p}{r_a+r_p},\,\,\, p=\frac{2r_a r_p}{r_a+r_p},\,\,\, \iota=\frac \pi 2 -\theta_{min}
\end{aligned}
\end{equation} 
where $r_a$ is apastron, $r_p$ is periastron and $\theta_{min}$ is the minimum of $\theta$ coordinate along the geodesics. In Kerr spacetime, we can determine the three orbit parameters $e,p,\iota$ from three conserved quantities $E,L_z,Q$ and vice versa. In KRZ non-Kerr spacetime, for equatorial orbits, we can still determine $e,p$ by $E,L_z$.

Transform $(r,\,\theta,\,\phi)$ into $(x,\,y,\,z)$ with the definition of spherical coordinates (rather than the Boyer-Lindquist coordinates), namely $x=r\sin\theta \cos \phi,\, y=r\sin\theta\sin\phi,\, z=r\cos\theta$. Then use quadrupole formula, i.e.
\begin{equation}
	\bar{h}^{jk}(t,{\rm {x}})= \frac 2 r [\ddot{T}^{jk} (t')]_{t'=t-r}
\end{equation}
\begin{equation}
	I^{jk}=\mu x'^j_p x'^k_p
\end{equation}
where $\bar{h}^{\mu\nu} = h^{\mu\nu} - \frac 1 2 \eta^{\mu\nu} \eta^{\rho\sigma} h_{\rho\sigma} $ is the metric perturbation under trace-reversed gauge

transform the waveform into transverse-traceless gauge (see formula (17) and (23) in \cite{kludge}), and we get the plus and cross component of the waveform observed at latitudinal angle $\Theta $ and azimuthal angle $\Phi$:
\begin{equation}
\begin{aligned}
h_+ =& h^{\Theta\Theta} - h^{\Phi\Phi}\\
=& \{ \cos^2 \Theta [ h^{xx} \cos^2 \Phi + h^{xy} \sin 2\Phi h^{yy} \sin^2 \Phi ] + h^{zz} \sin^2\Theta - \sin 2\Theta [h^{xz} \cos \Phi + h^{yz} \cos \Phi ] \}\\
&- [ h^{xx} \sin^2\Phi - h^{xy} \sin 2\Phi + h^{yy} \cos^2 \Phi ]
\end{aligned}
\end{equation}\\
\begin{equation}
\begin{aligned}
h_\times =& 2 h^{\Theta\Phi} \\
=& 2 \{ \cos\Theta[-\frac 12 h^{xx} \sin 2\Phi + h^{xy} \cos 2\Phi + \frac 12 h^{yy} \sin 2\Phi ] + \sin \Theta [h^{xz} \sin \Phi - h^{yz} \cos \Phi ] \}
\end{aligned}
\end{equation}
With the resulted ``plus" and ``cross" components, we define our waveforms as $h = h_+ + i h_\times$

Matched filtering is the standard technique to be used in LISA analysis. In real EMRI data analysis, a large bank of waveform templates will be compared with the detected signal to find the matched template. Here we mainly adopt the fitting factor as a measure of similarity between two waveforms within LISA band. 

The inner product between a signal, $a(t)$, and a template, $b(t)$, is defined by their cross correlation: \cite{product}
\begin{equation}
	(a|b)=4\Re\int \frac{\tilde{a}^*(f) \tilde{b}(f)}{S_n(f)}df =2\int \frac{\tilde{a}^*(f) \tilde{b}(f) +\tilde{a}(f) \tilde{b}^*(f) }{S_n(f)}df
\end{equation}
where $S_n(f)$ is the power spectral density of LISA noise. In our calculation, the analytic fit to the noise spectrum same as \cite{kludge} is used.

The overlap (fitting factor) between the signal and template is defined as:
\begin{equation}
	\rm {FF}(a,b)=\frac{(a|b)}{\sqrt{(a|a)(b|b)}}
\end{equation}
When the overlap between two waveforms is above 0.97, we believe the waveform of this template is discovered in the signal. However, if a non-Kerr signal has an overlap above 0.97 with a Kerr template, we could mistake the signal as emitted from around a Kerr BH, i.e. the confusion problem mentioned in \cite{sameOmg}.

\section{Numerical Results and Analysis}
\label{p_conf}
When we try to identify EMRI signals, the confusion problem, as described in \cite{sameOmg}, could prevent us from discerning non-Kerr signal and Kerr signal. Namely an overlap over 0.97 might exist between non-Kerr signals and Kerr ones of certain parameters. In \ref{p_2d} and \ref{p_3d} we show the confusion problem when matching EMRIs from equatorial and inclined orbit.
\subsection{Equatorial orbit}
\label{p_2d}
Given a waveform under spacetime with non-zero deformation, in order to see if there is ``confusion problem", searching over the entire parameter space would be computationally impossible. A better way is to have some idea about which waveform under Kerr spacetime is most similar to the non-Kerr signal and look at their overlap. Here we search for existence confusion problem with similar method as \cite{majorPRD}, i.e. looking at waveforms generated from geodesics with same orbital frequencies. The orbital frequencies in Kerr spacetime are given in \cite{tauOmg}. In equatorial orbits, there are two frequencies $\omega_\phi$ and $\omega_r$ related to motion of $\phi$ and $r$ coordinates.

\begin{figure}[!ht]
	\centering
	\includegraphics[width=16cm]{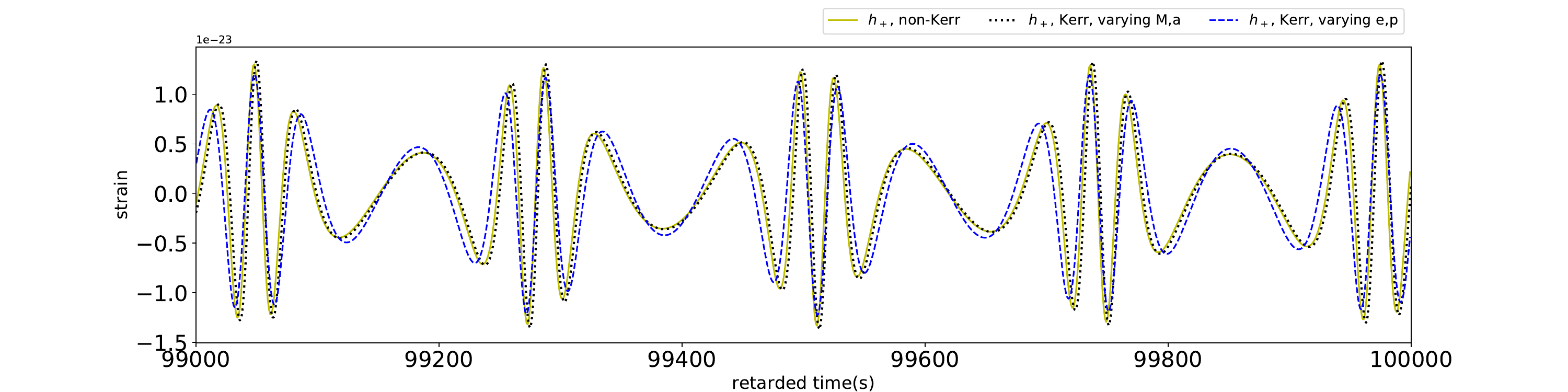}
	\includegraphics[width=16cm]{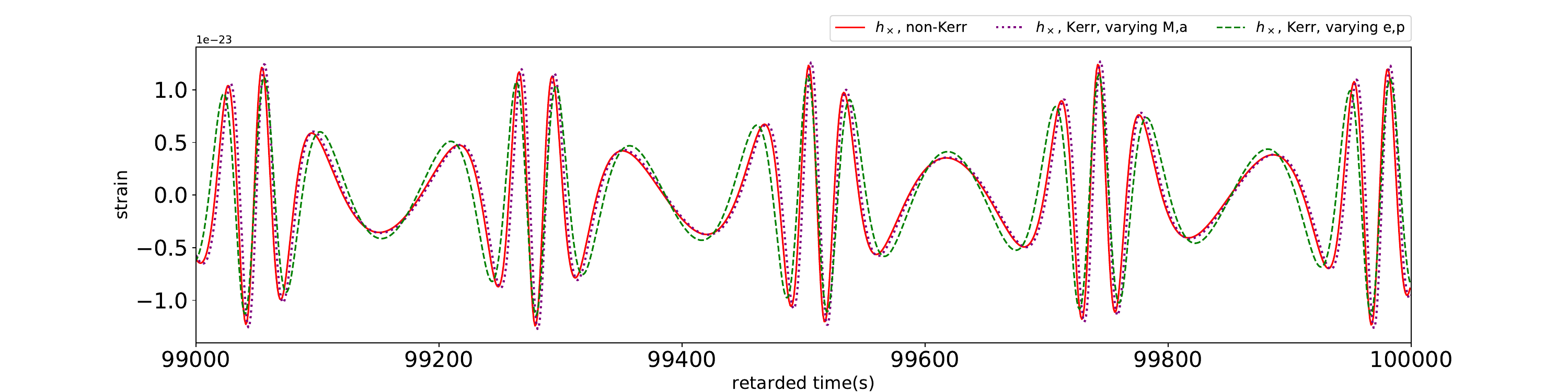}
	\caption{``Plus" and ``cross" component of EMRIs under non-Kerr spacetime and Kerr spacetime varying $(M,a)$ or $(e,p)$ to equate orbital frequency. The non-Kerr waveform is $(\delta_1,a,M,e,p)=(0,2,0.5,2\times10^5, 0.5,6.0)$. The yellow and red solid lines are ``plus" and ``cross" component of the non-Kerr waveforms. The black and violet dotted lines are `plus" and ``cross" component of Kerr waveforms with same orbital frequencies as non-Kerr orbit by varying $(M,a)$. The blue and green dashed lines are ``plus" and ``cross" component of Kerr waveforms with same orbital frequencies as non-Kerr orbit by varying $(e,p)$.}
	\label{kkwave}
\end{figure}	

For equatorial motions, we set the initial $t$ and $\phi$ to 0 in view of stationarity and axisymmetry and set initial $r=r_{max}$, so the orbit is uniquely determined by orbital eccentricity $e$, semilatus rectum $p$, deformation parameters $\delta_i$, BH mass $M$ and BH spin $a$. As described in \cite{majorPRD}, we can achieve same orbital frequency as non-Kerr orbits by varying orbital parameters $e, \,p$ or BH parameters $M, \, a$. So we need to consider EMRIs determined by $(\delta,\, a,\, M,\, e,\, p)$, $(0,\, a,\, M,\, e_{\rm{Kerr}},\, p_{\rm{Kerr}})$ and $(0,\, a_{\rm{Kerr}},\, M_{\rm{Kerr}},\, e,\, p)$. Comparison of waveforms generated by orbits of same orbital frequency is show in Fig. \ref{kkwave}. The overlap between waveforms varying BH mass and spin is over 0.99. In fact, the geodesics that generate the two waves are overlapping.

\begin{figure}[htb]
	\centering
	\includegraphics[width=7cm]{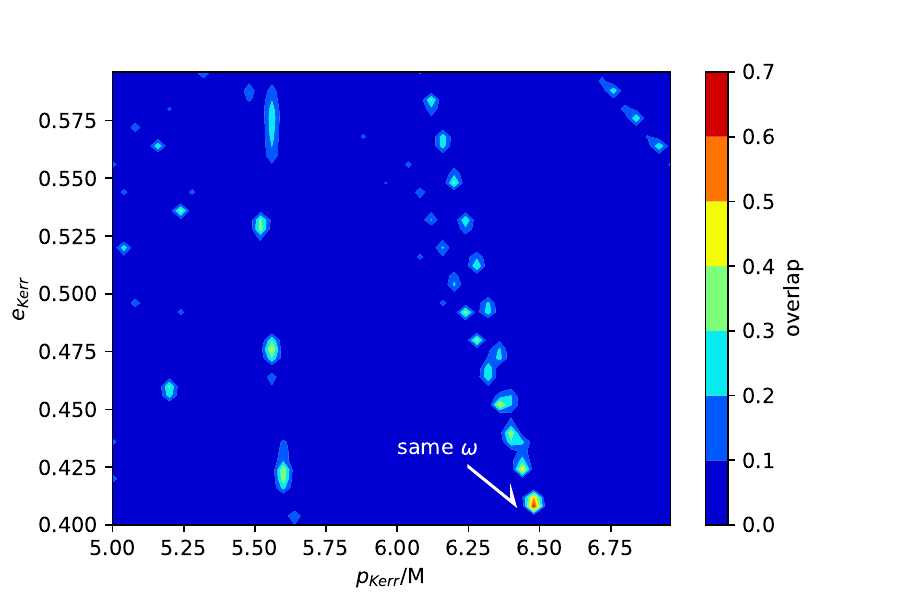}
	\includegraphics[width=7cm]{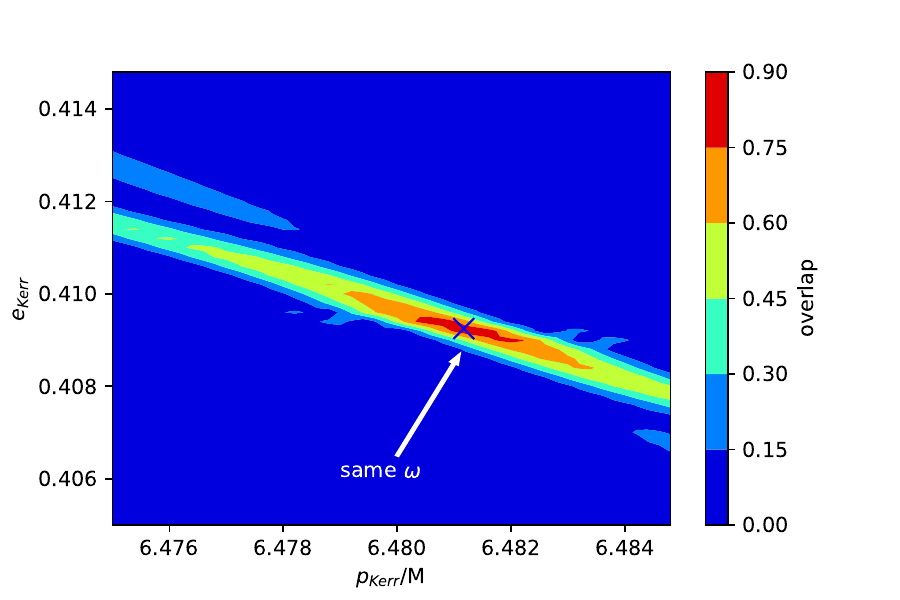}
	
	\caption{Distribution of overlap between waveforms defined by ($\delta_1,\, a,\, M,\, e,\, p$) = (0.2, 0.5, $2 \times 10^5 $, 0.5, 6) and  ($\delta_1,\, a,\, M,\, e,\, p$) =(0, 0.5, $2 \times 10^5 $, $e_{\rm{Kerr}}$, $p_{\rm{Kerr}}$) on ($e_{\rm{Kerr}}$, $p_{\rm{Kerr}}$) plane. Blue cross mark pointed by the arrow: same $\omega_r$ and $\omega_\phi$ at ($e_{\rm{Kerr}}$, $p_{\rm{Kerr}}$) = (0.409248, 6.481170).  The grid size is 50*50.}
	\label{overlapdist}
\end{figure}

According to Ref. \cite{sameOmg}, orbits with same orbital frequency $\omega_r$ and $\omega_\phi$ can generate gravitational waveforms potentially confused with non-Kerr signals. Therefore the overlap between a non-Kerr waveform and several Kerr waveforms should have a local maximum around the parameter leading to same orbital frequency. Here we check this result by looking at overlaps between waveforms defined by ($\delta_1,\, a,\, M,\, e,\, p$) = (0.2, 0.5, $2 \times 10^5 $ , 0.5, 6) and of ($\delta_1,\, a,\, M,\, e,\, p$) = (0, 0.5, $2 \times 10^5 $ , $e_{\rm{Kerr}}$, $p_{\rm{Kerr}}$) with varied $e_{\rm{Kerr}}$ and $p_{\rm{Kerr}}$. First we looked at overlap distribution on a relatively large range of (e, p) and found the highest local maximum locates around the point whose parameters corresponds to identical orbital frequencies. Then we searched near ($e_{\rm{Kerr}}$, $p_{\rm{Kerr}}$) with same orbital frequency and found the the aforementioned point approximately locates at the peak of overlap distribution , as shown in Fig. \ref{overlapdist}. 

\begin{figure}[!h]
	\centering
	\includegraphics[width=7.4cm]{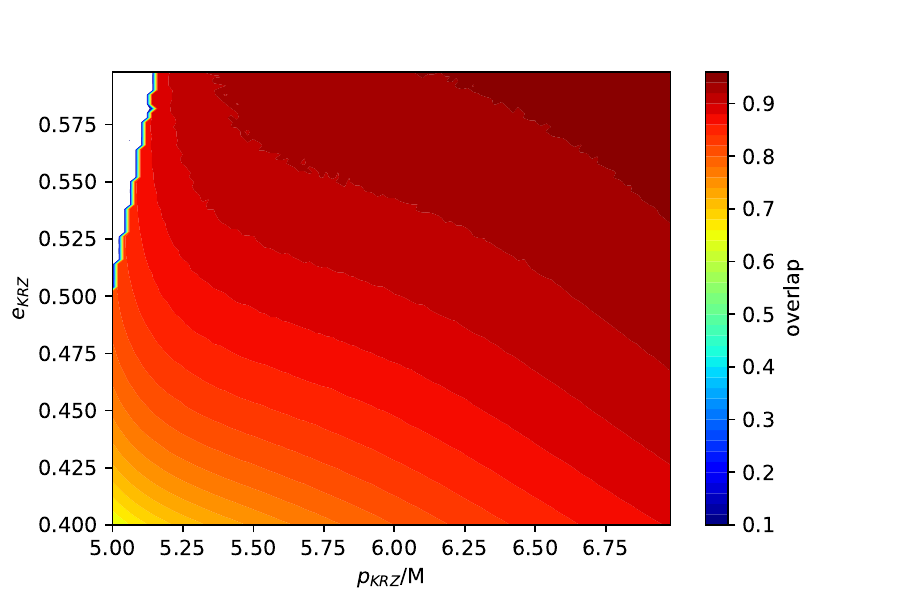}
	\includegraphics[width=7.4cm]{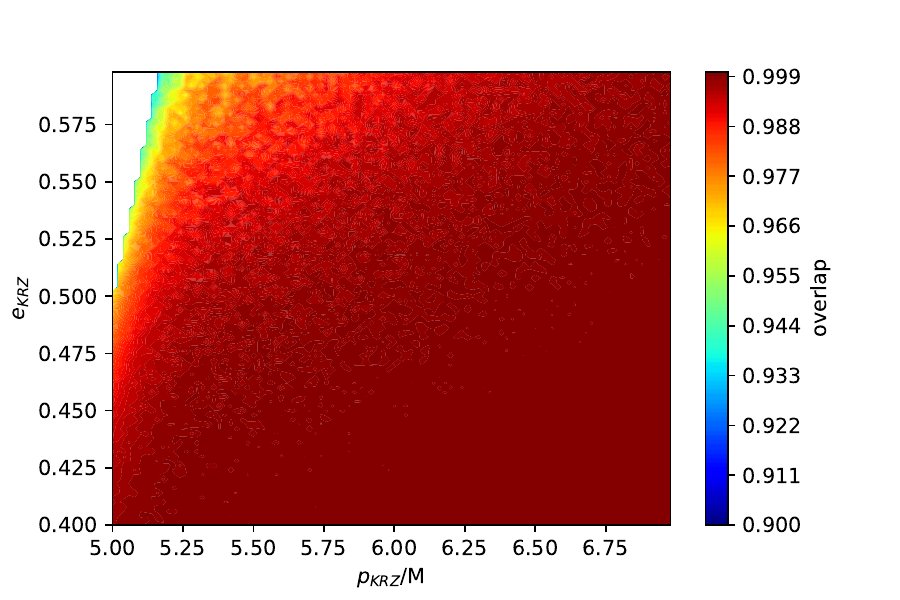}
	\caption{distribution of overlap between waveforms of ($\delta_1,\, a,\, M,\, e,\, p$) = (0.2, 0.5, $2 \times 10^5 $ , $e_{\rm{KRZ}}$, $p_{\rm{KRZ}}$) and Kerr waveforms with identical $\omega_r$ and $\omega_\phi$ by varying $e_{\rm{Kerr}},\, p_{\rm{Kerr}}$ (left panel) and varying $M,a$ (right panel), on $(e_{\rm KRZ},p_{\rm KRZ})$ plane. The grid size is 100*100. The blank region is unstable orbits.}
	\label{epdist}
\end{figure}	

%
%

Then we investigated the confusion problems for different non-Kerr signals. First we studied the signals from different orbits, i.e. different $e_{\rm KRZ}, \, p_{\rm KRZ}$. Since metric deformation is more evident near BH horizon, we look at waveforms generated by trajectories close to innermost bound orbit. We compare waveforms defined by ($\delta_1,\, a,\, M,\, e,\, p$) = (0.2, 0.5, $2 \times 10^5 $ , $e_{\rm KRZ}$, $p_{\rm KRZ}$) and Kerr orbits with same orbital frequencies by varying $(e,p)$ or $(M,a)$ in a 100*100 grid of $(e,p)$. Contour plots of waveform overlap when varying $(e,p)$ and $(M,a)$ are shown in Fig. \ref{epdist}. \comment{and \ref{amdist}.} The confusion problem exists when varying $(M,a)$ for most parameter region we considered. 

\begin{figure}[!ht]
	\centering
	\includegraphics[width=7.4cm]{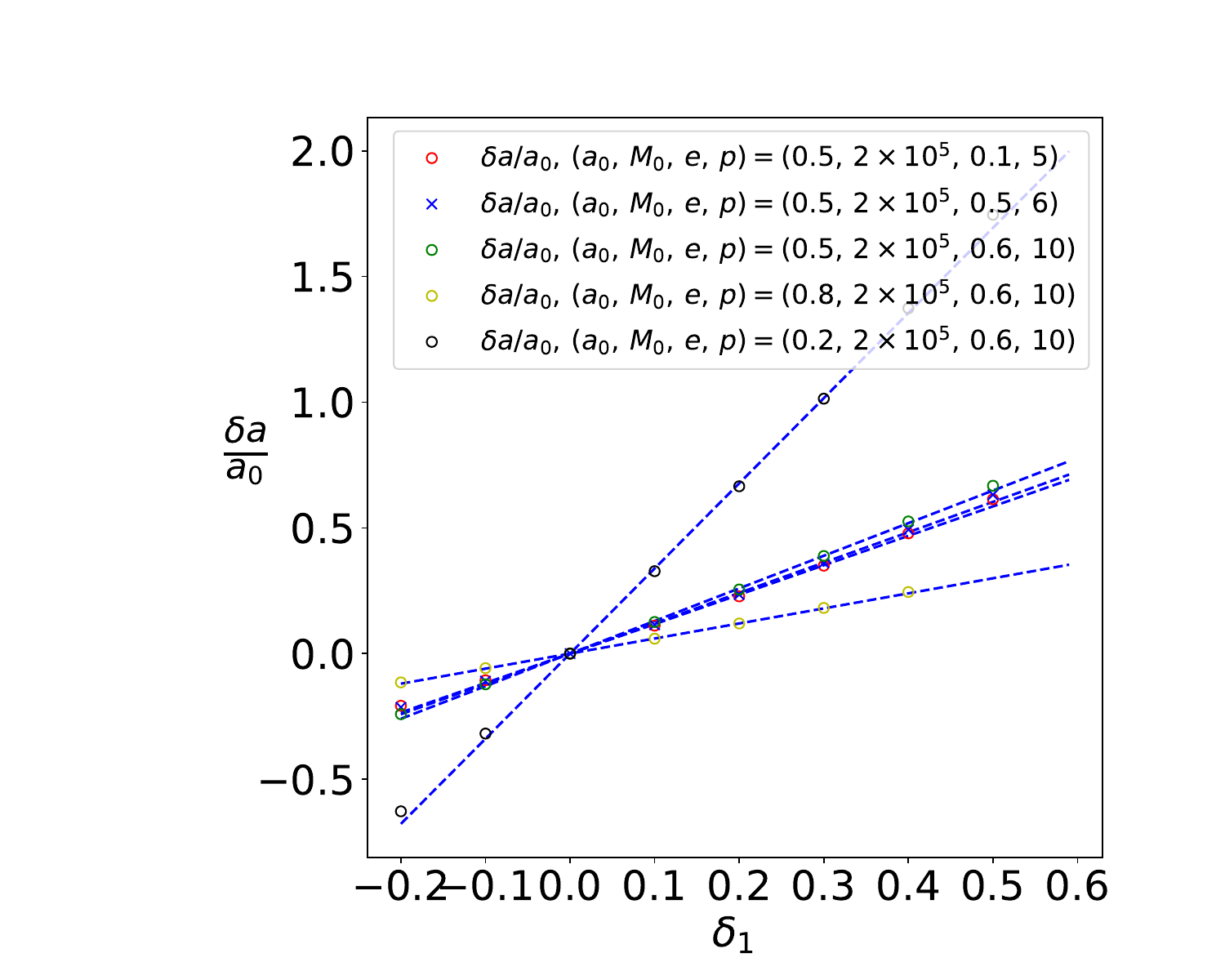}
	\includegraphics[width=7.4cm]{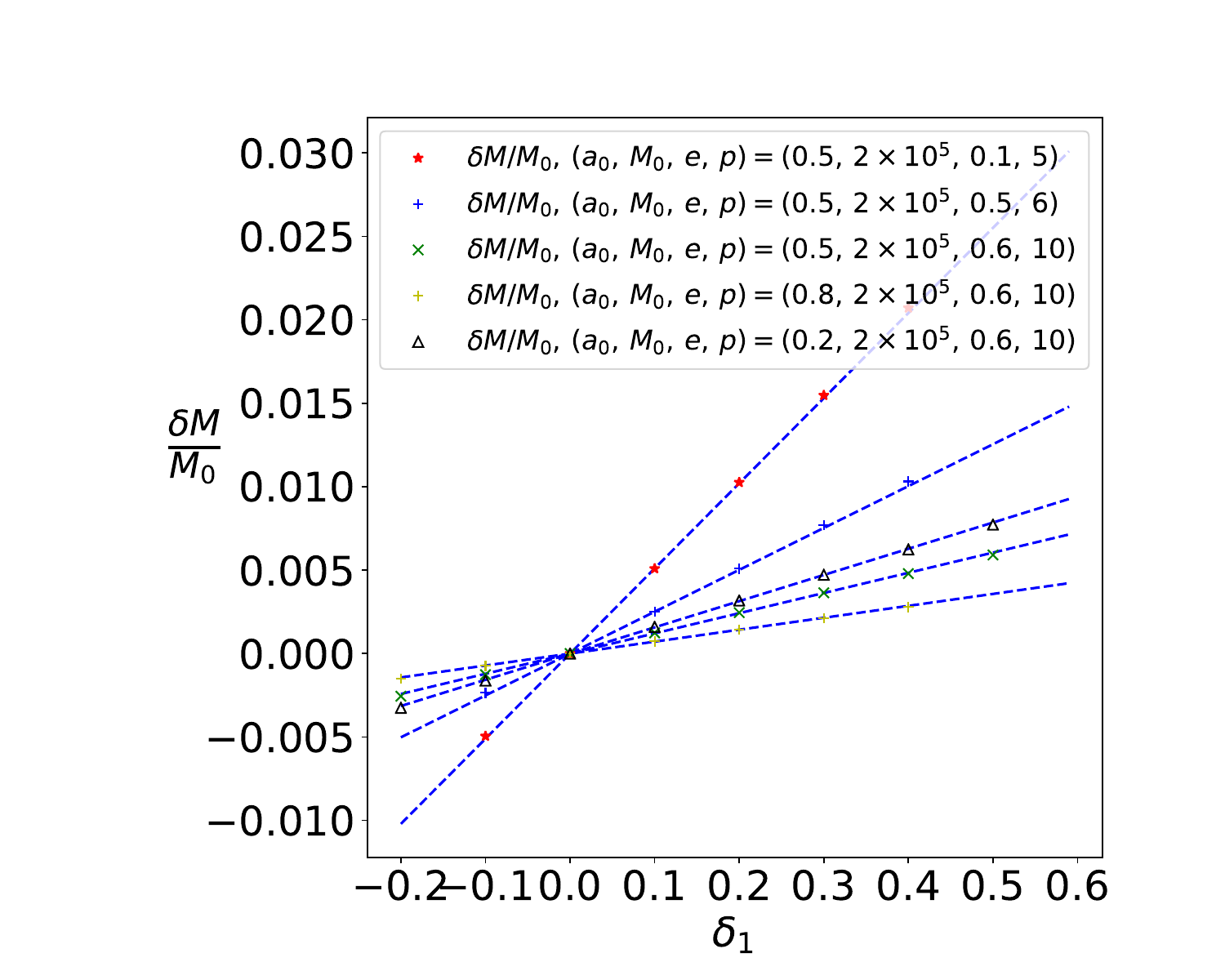}
	\caption{relation between relative varied spin/Mass when equating orbital frequency, i.e. orbit of $(\delta_1, a_0, M_0,e,p)$ and $(0,a_0+\delta a, M_0+\delta M,e,p)$ have same orbital frequencies and here shows $\frac{\delta a}{a_0}$-$\delta_1$ and $\frac{\delta M}{M_0}$-$\delta_1$ relations. Left panel: relations of $\frac{\delta a}{a_0}$ to $\delta_1$, Right panel: relations of $\frac{\delta M}{M_0}$ to $\delta_1$.}
	\label{da_linear}
\end{figure}

\begin{figure}[!ht]
	\centering
	\includegraphics[width=7.4cm]{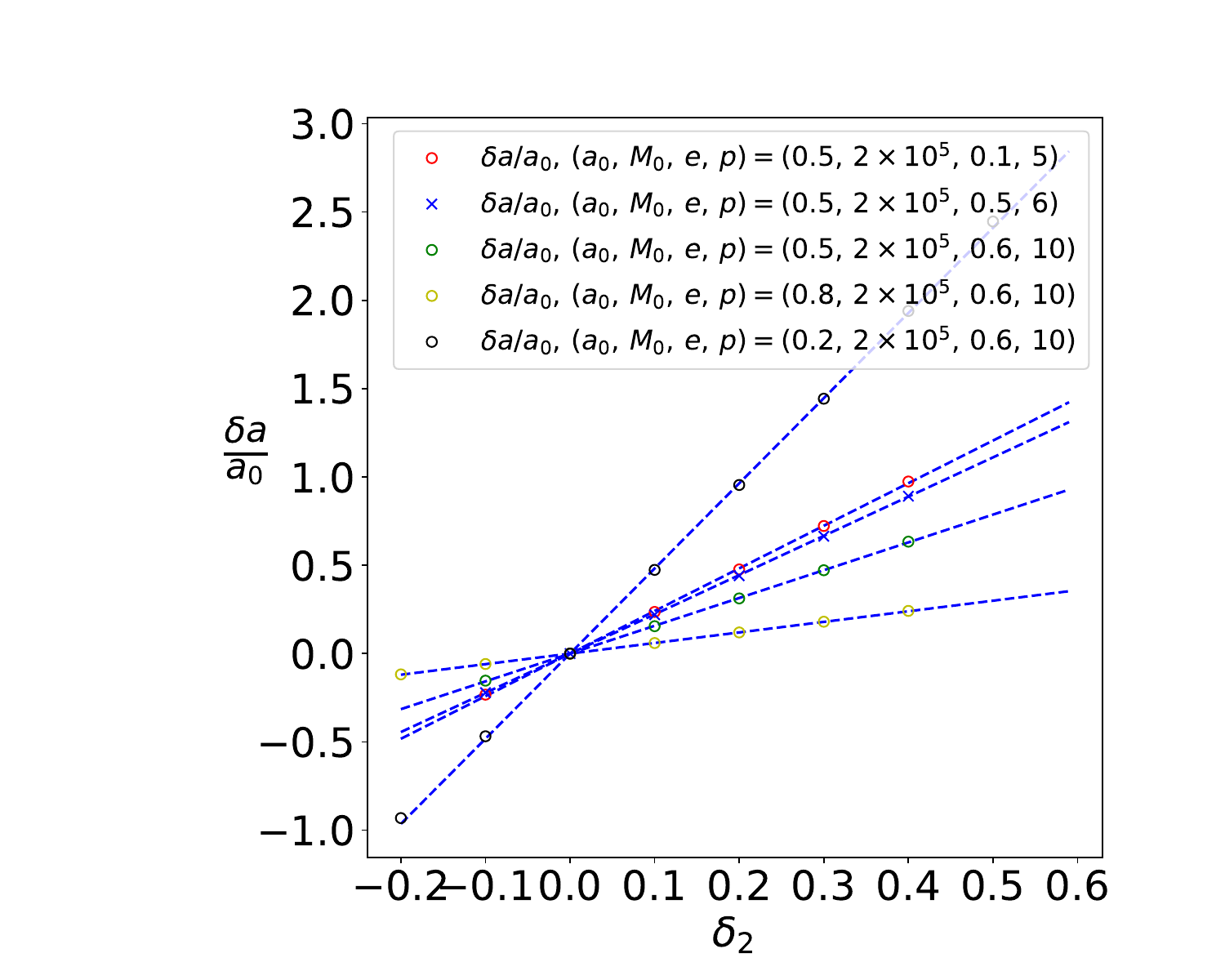}
	\includegraphics[width=7.4cm]{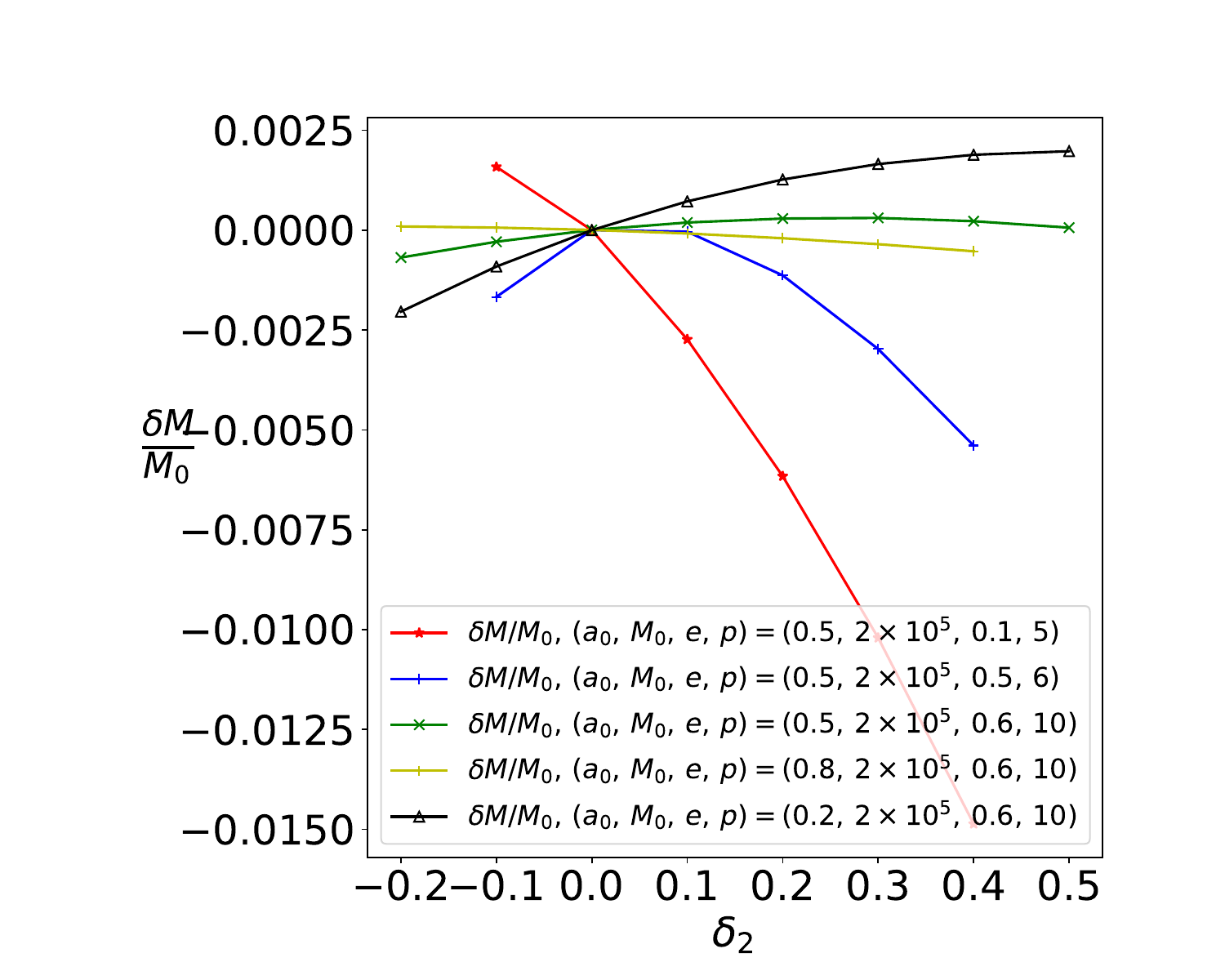}
	\includegraphics[width=7.4cm]{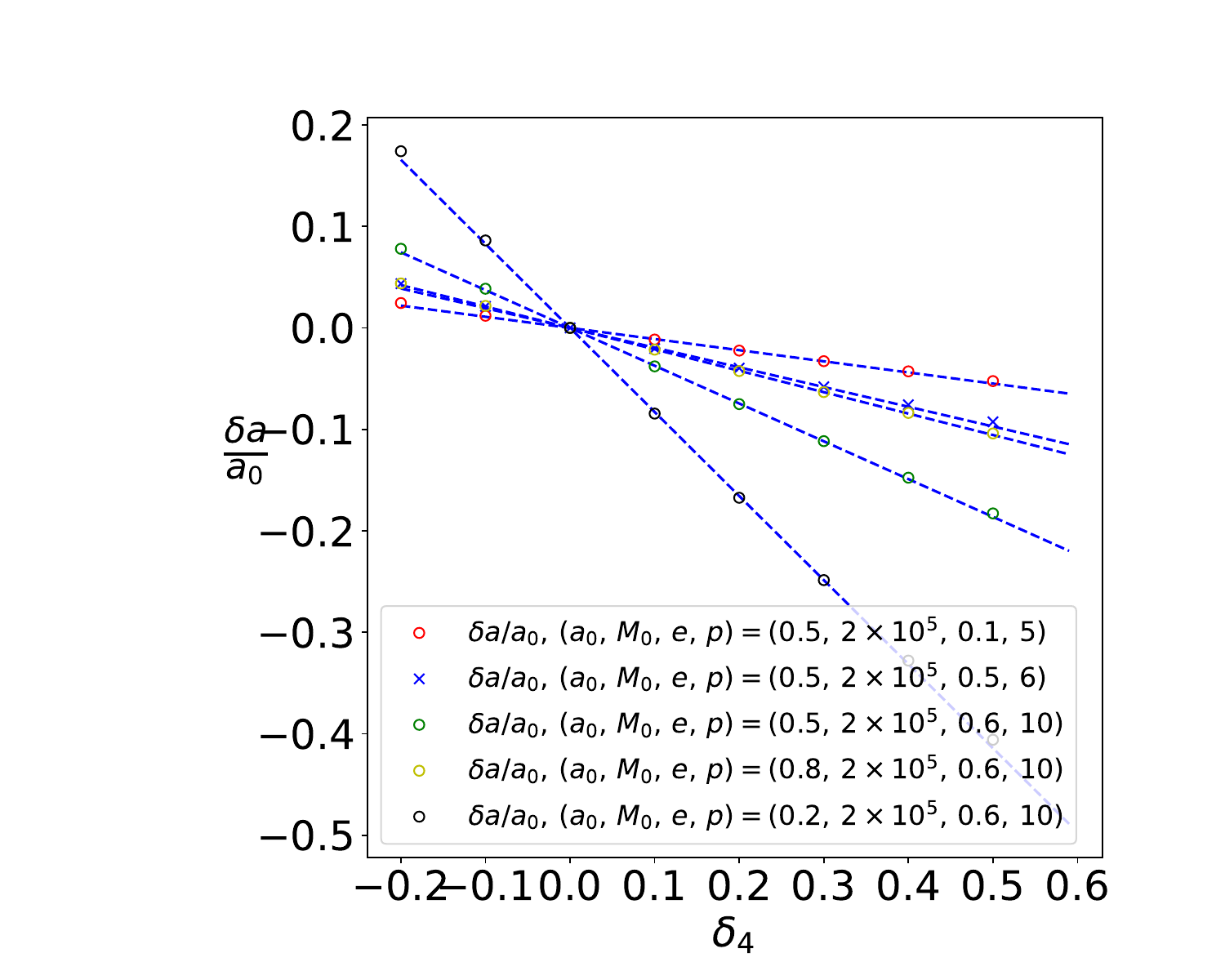}
	\includegraphics[width=7.4cm]{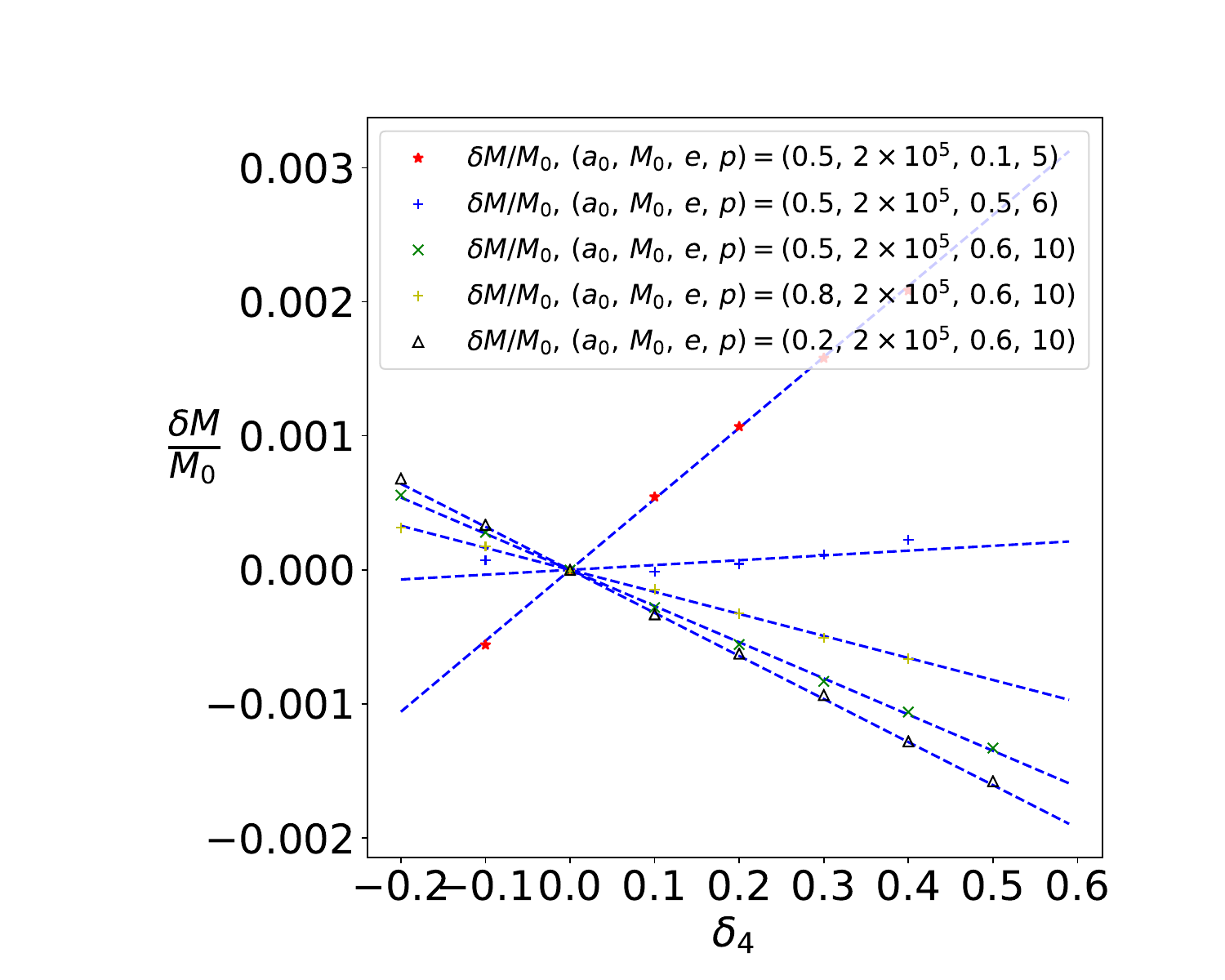}
	\caption{The same as Fig. \ref{da_linear}, but for $\delta_2$ (top panels) and $\delta_4$ (bottom panels).}
	\label{d2_linear}
\end{figure}


\begin{figure}[!ht]
	\centering
	\includegraphics[width=8cm]{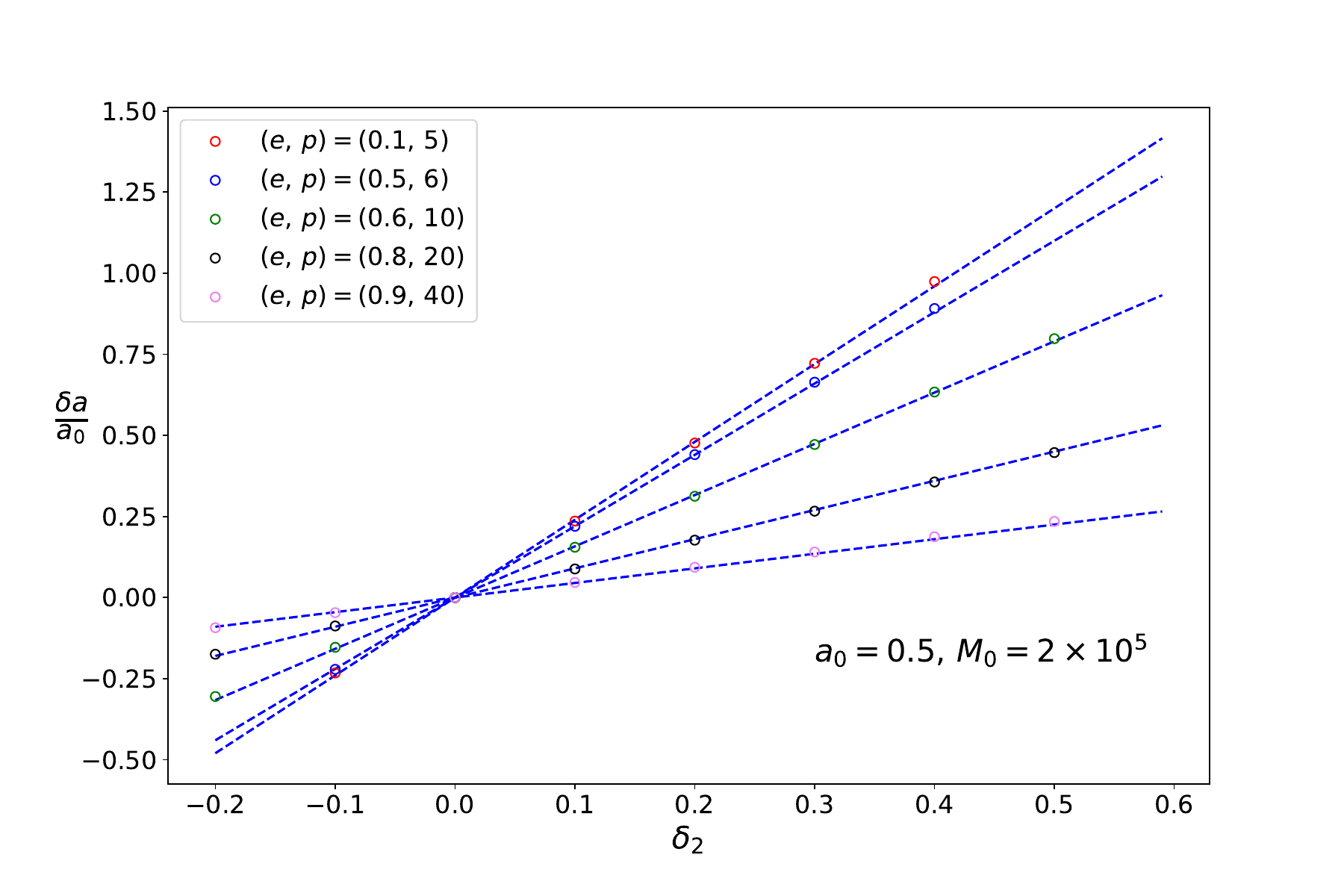}
	
	\caption{The same as Fig. \ref{da_linear}, but only for $\delta a/a_0$-$\delta_2$ relations for different orbits.}
	\label{ep_slope}
\end{figure}

Furthermore, the confusion problem exists for a large range of deformation parameter. On top of that, when changing the deformation parameter, we found an almost linear relation between $\delta_i$ and varied BH spin/mass, $a_{\rm{Kerr}}$ and $M_{\rm{Kerr}}$, as shown in Fig. \ref{da_linear}, \ref{d2_linear}. \comment{and \ref{d4_linear}.} Since we are considering equatorial orbits i.e. $\cos\theta=0$, only $\delta_1$, $\delta_2$ and $\delta_4$ have influences. Fig. \ref{ep_slope} shows that slope of this ``linear" relation varies greatly for different $(e,\,p)$, so the relation is not a property intrinsic to the metric but dependent on the orbit. The linearity is not a result of small deformation. In fact, as shown in the figures above, the spin varies up to one or two times the spin in KRZ metric. The linearity does not hold for $\delta M-\delta_2$, but since it's not our major concern and the mass here is just a time scale, we did not dig further into it. 

\begin{figure}[!ht]
	\centering
	\includegraphics[width=7cm]{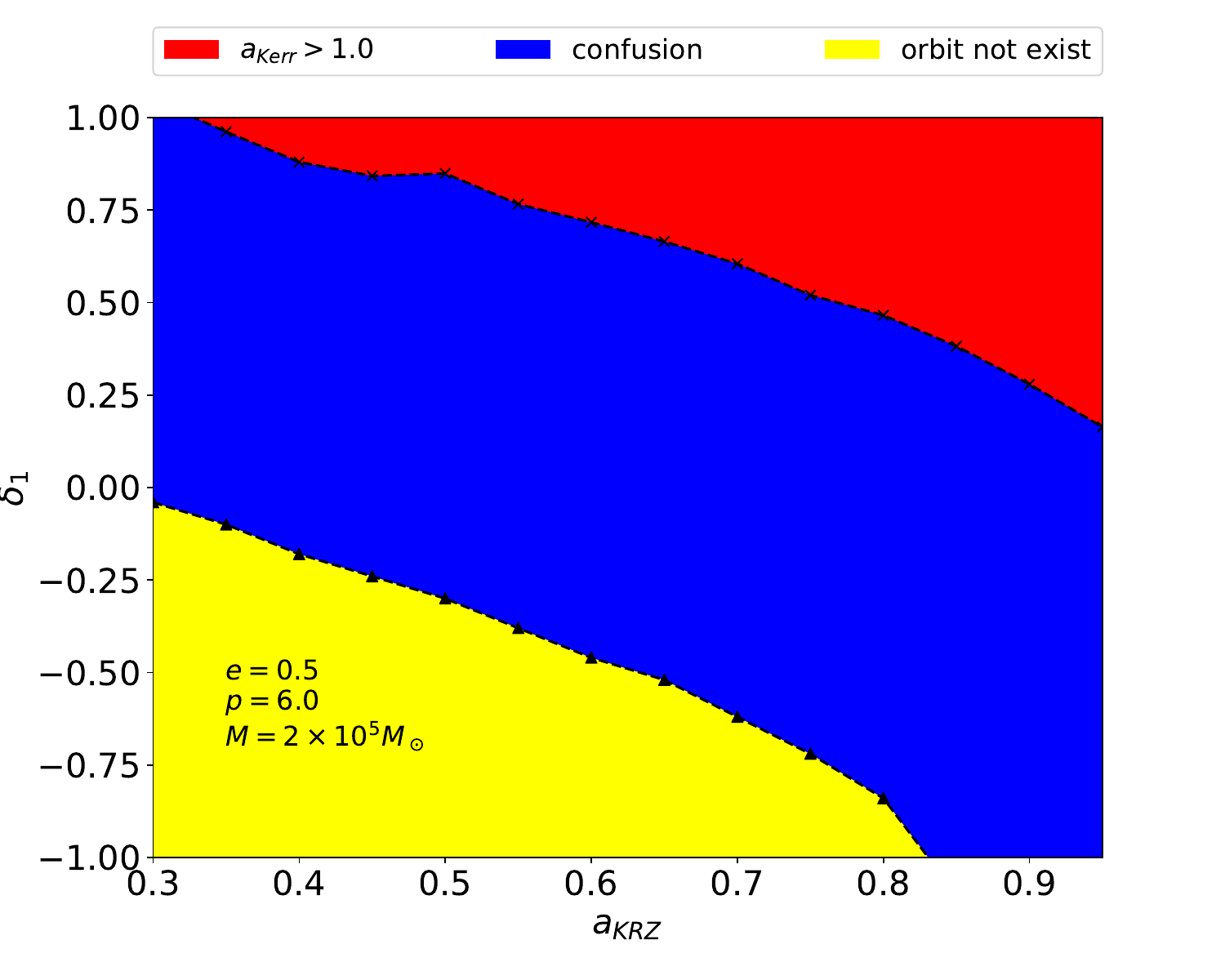}
	\includegraphics[width=7cm]{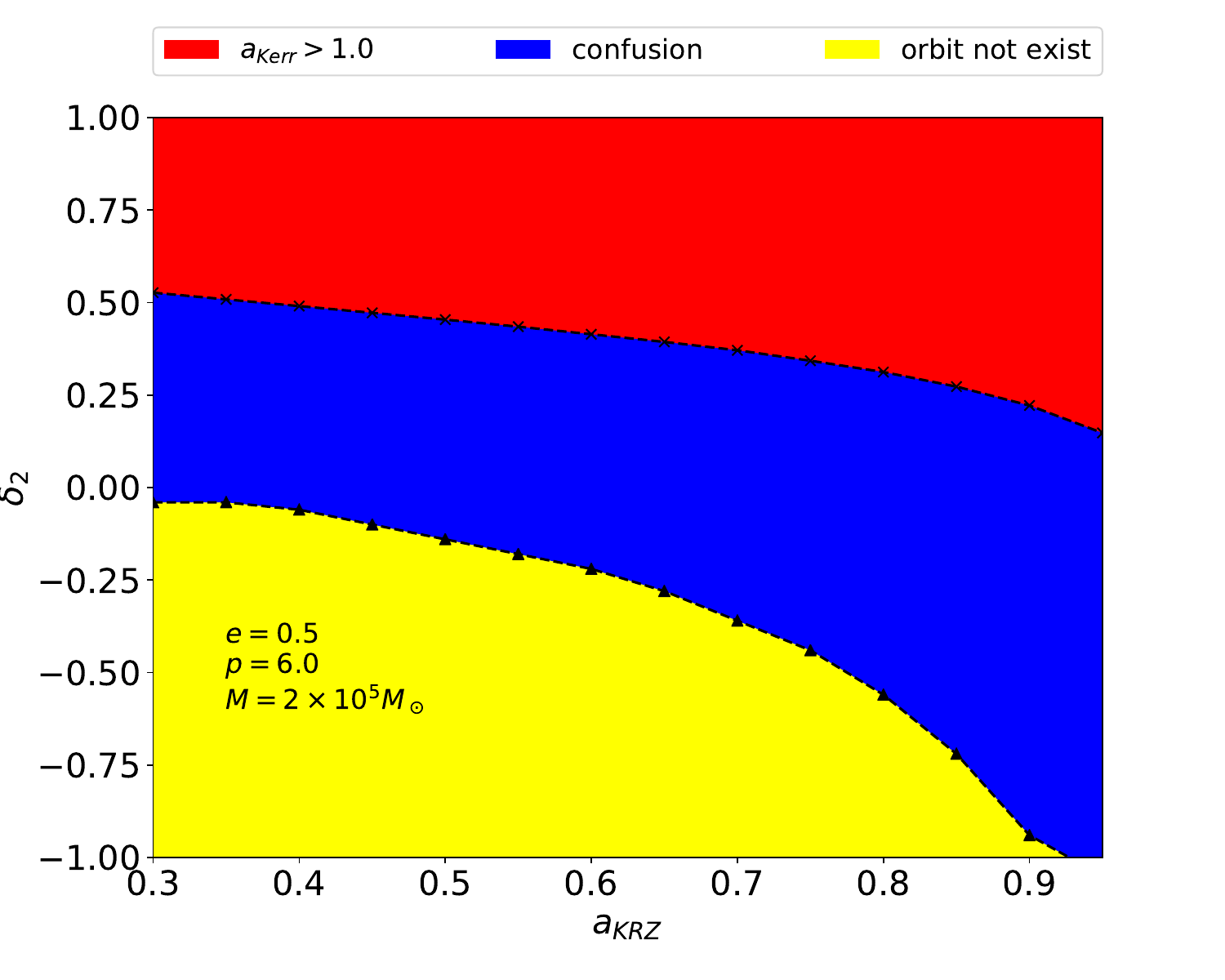}
	
	\caption{Upper and lower limit of $\delta_i$, whose signals can be confused with Kerr waveform, on $delta_i$-$a_{\rm{KRZ}}$ parameter plane, set by equatorial orbit requiring $a_{\rm{Kerr}}<1$ and stable orbit. Other parameters are set to: eccentricity $e=0.5$, semi-latus $p=6.0$ and Balck hole mass $M=2\times10^5$ solar mass. Waveforms in red parameter region, when trying to equate orbital frequencies by varying $(M,a)$, results in $a_{\rm{Kerr}}$ larger than 1. Waveforms in blue region can be confused with Kerr waveform. Orbits in yellow region do not exist. Left panel: for $\delta_1$ and $\delta_2$ set to 0. Right panel: for $\delta_2$ and $\delta_1$ set to 0. }
	\label{d2limit}
\end{figure}

Therefore, for a given orbit parameter $(e,\, p)$, we can regard the introduction of $\delta_i$ as adding the black hole spin and mass proportionally. This sets an limit for the range of deformation parameters within which we can play the trick of varying $(M_{\rm{Kerr}},\, a_{\rm{Kerr}})$. The upper limit is set by requiring $a_{\rm{Kerr}}<1$ and the lower is limited by $a_{\rm{Kerr}}>0$ or stable orbit, e.g. for $(e,\, p)=(0.5,\, 6.0)$, when $a_{\rm{Kerr}}$ is small, the orbit is no longer stable and bounded. Fig. \ref{d2limit} shows the upper and lower bound of deformation parameters with respect to BH spin for orbit with $e=0.5,\, p=6.0$. 

\subsection{Inclined orbit}
\label{p_3d}
Equatorial orbits set some special conditions, i.e. number of orbital frequencies is equal to number of Kerr BH parameters. Therefore, in general we can solve mass and spin by equating the two frequencies set by KRZ orbit, and the resulted geodesics are almost identical. However, astrophysical EMRIs are usually generated by inclined orbits, which have three orbital frequencies. In such cases we usually cannot equate the three frequencies by only varying BH parameters. 

\begin{figure}[H]
	\centering
	\includegraphics[width=7cm]{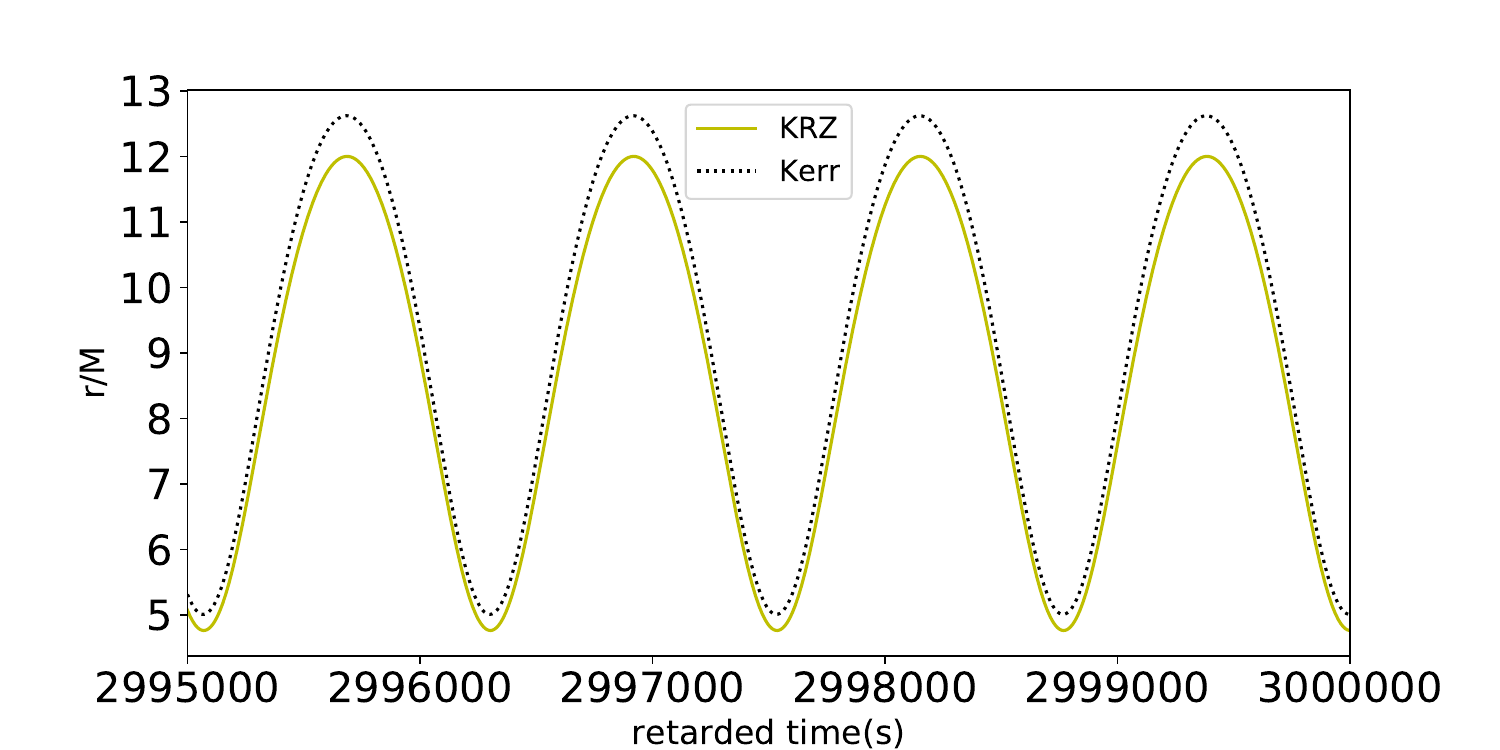}
	\includegraphics[width=7cm]{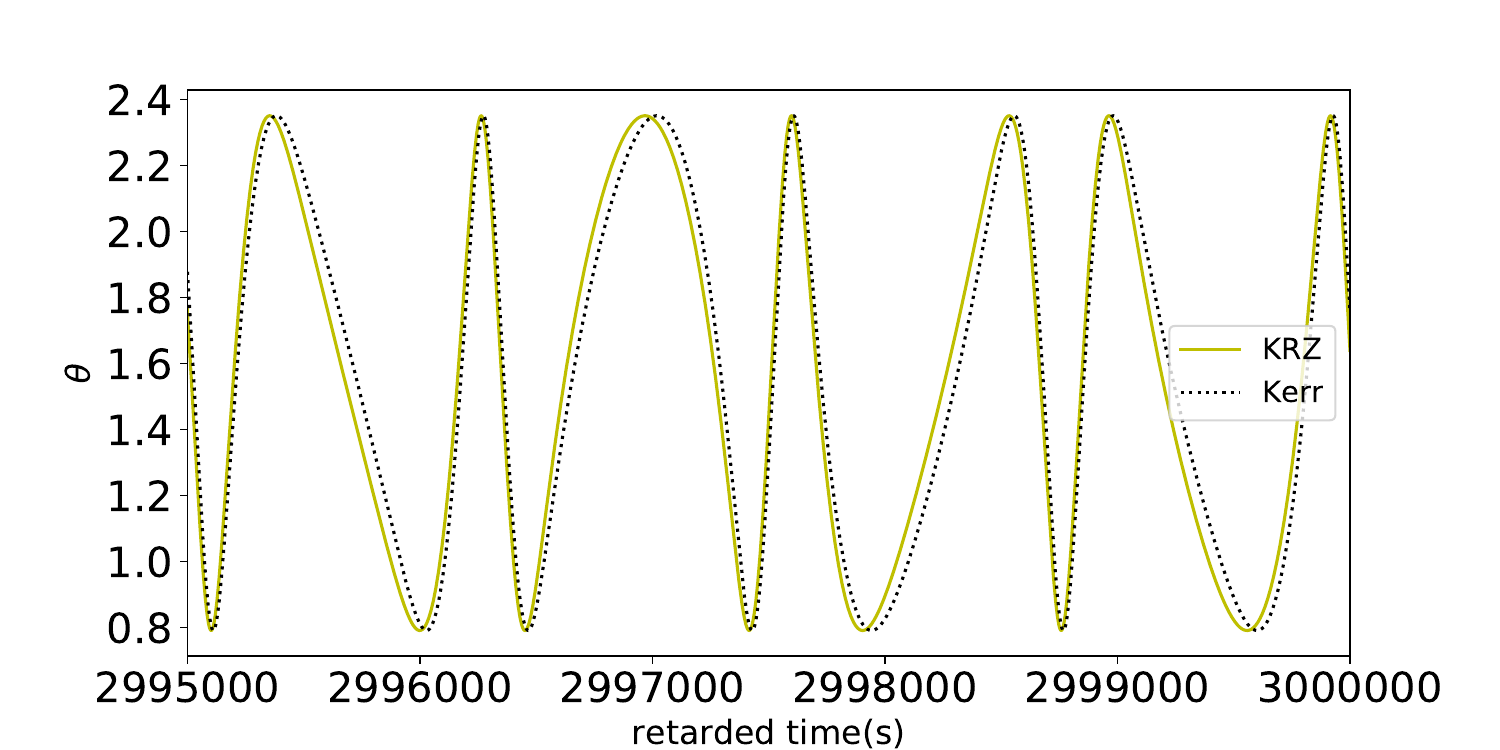}
	\includegraphics[width=7cm]{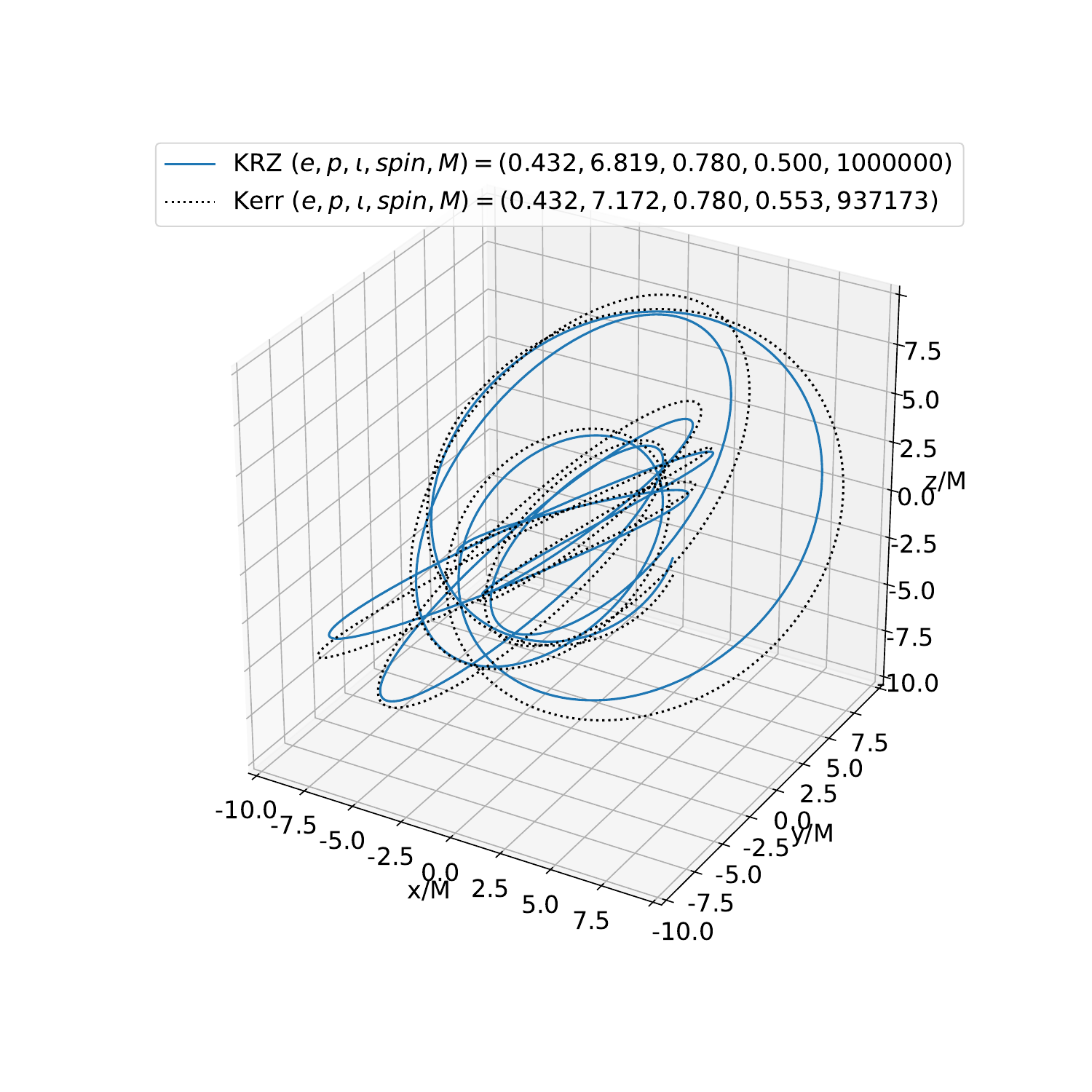}
	\includegraphics[width=7cm]{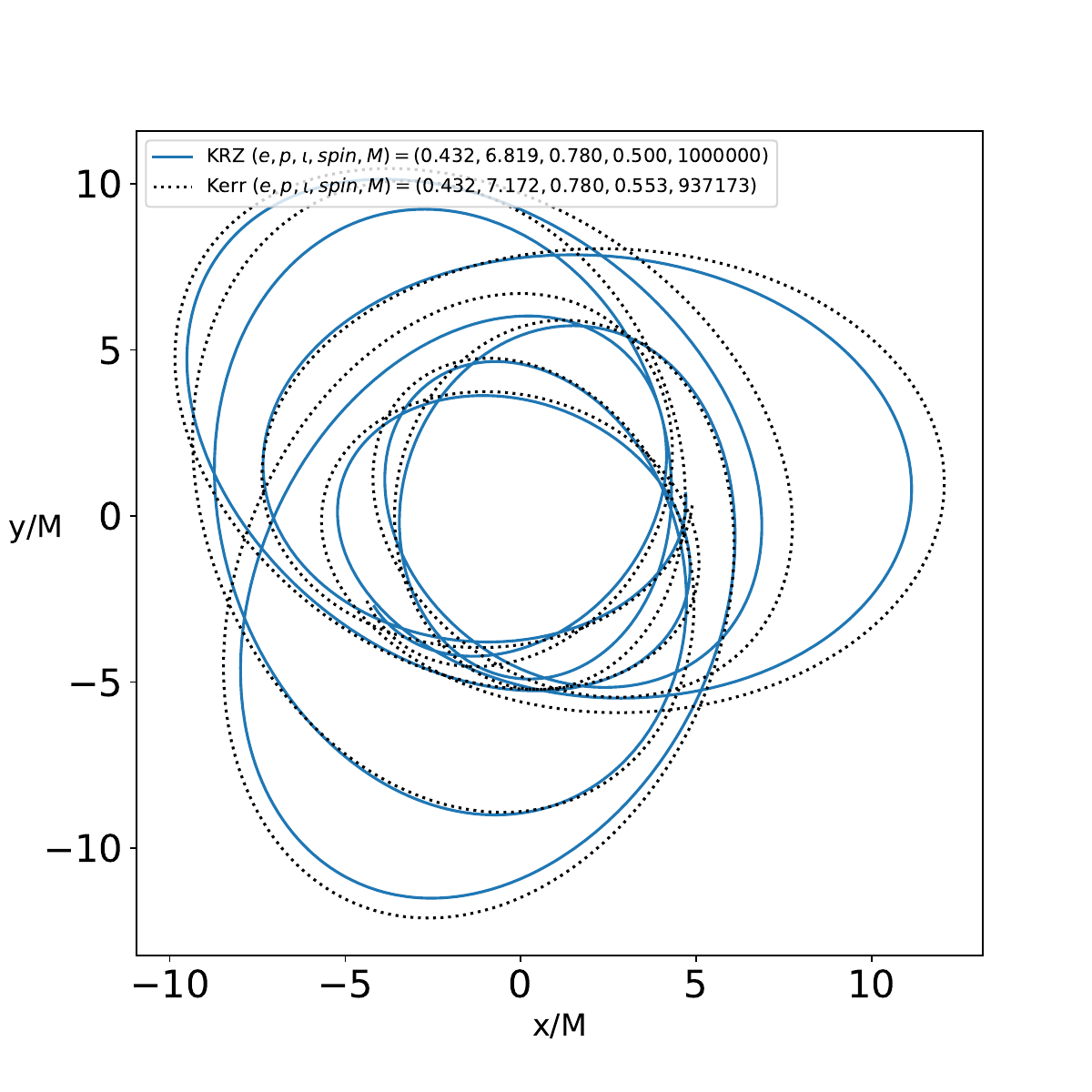}
	\includegraphics[width=14cm]{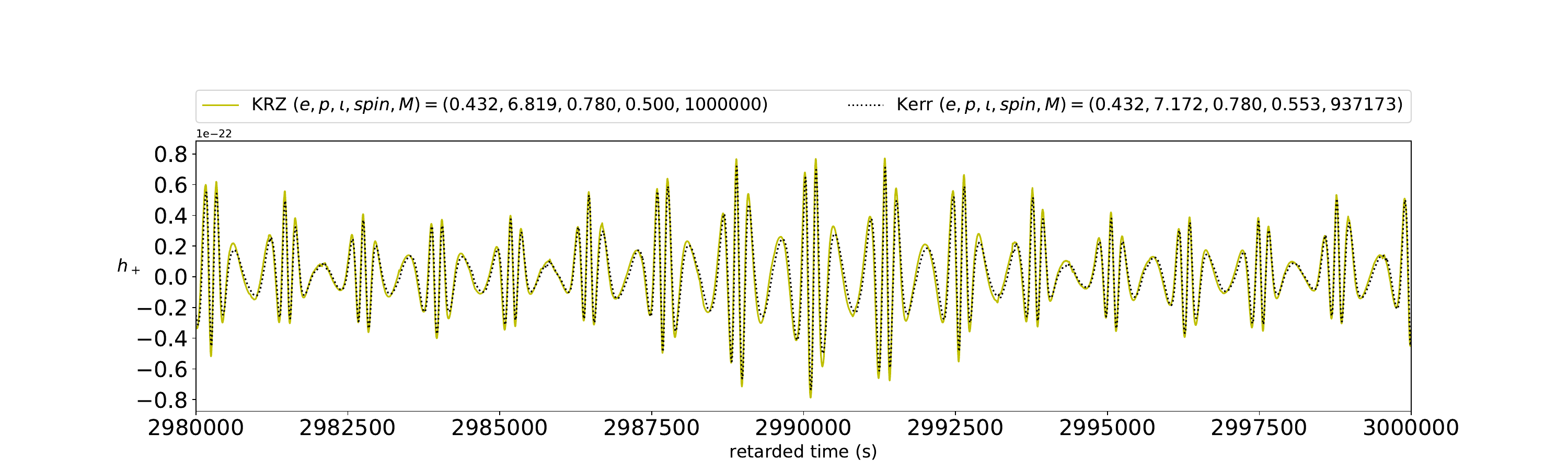}
	
	\caption{Orbits and GW waveforms of a KRZ orbit and a Kerr one with same orbital frequencies. Top left panel: time series of motion in r direction in last 5000s. Top right panel: time series of motion in $\theta$ direction in last 5000s. Middle left panel: trajectories in last 5000s. Middle right panel: projection on xy-plane of the trajectories in last 5000s. Bottom panel: ``plus" component of EMRI waveform in last 20000s. The orbit parameters and BH spin/Mass are $(e,p,\iota,spin,M)=(0.432,\, 6.819,\, 0.780,\, 0.5,\, 10^6)$. The deformation parameter of the KRZ orbit is $\delta_1=0.2$.}
	\label{3dtraj}
\end{figure}

However, we found that by varying $(M,\, a,\, p)$ to equate three orbital frequencies, the resulted gravitational waveforms also have an overlap over 0.97, even though the orbits are apparently not identical. Upper and middle panel of Fig. \ref{3dtraj} show the time series of motion in $r,\, \theta $ direction and trajectories in the last 5000s, the total time is $3\times 10^6$s. Motion at $\theta$ direction is almost overlapping but has a few distortions. Motion at $r$ direction has the same frequency but different amplitude, basically resulted from varying the semi-latus $p$. However, the gravitational wave signal are almost identical as shown in the bottom panel. 

\begin{figure}[H]
	\centering
	\includegraphics[width=8cm]{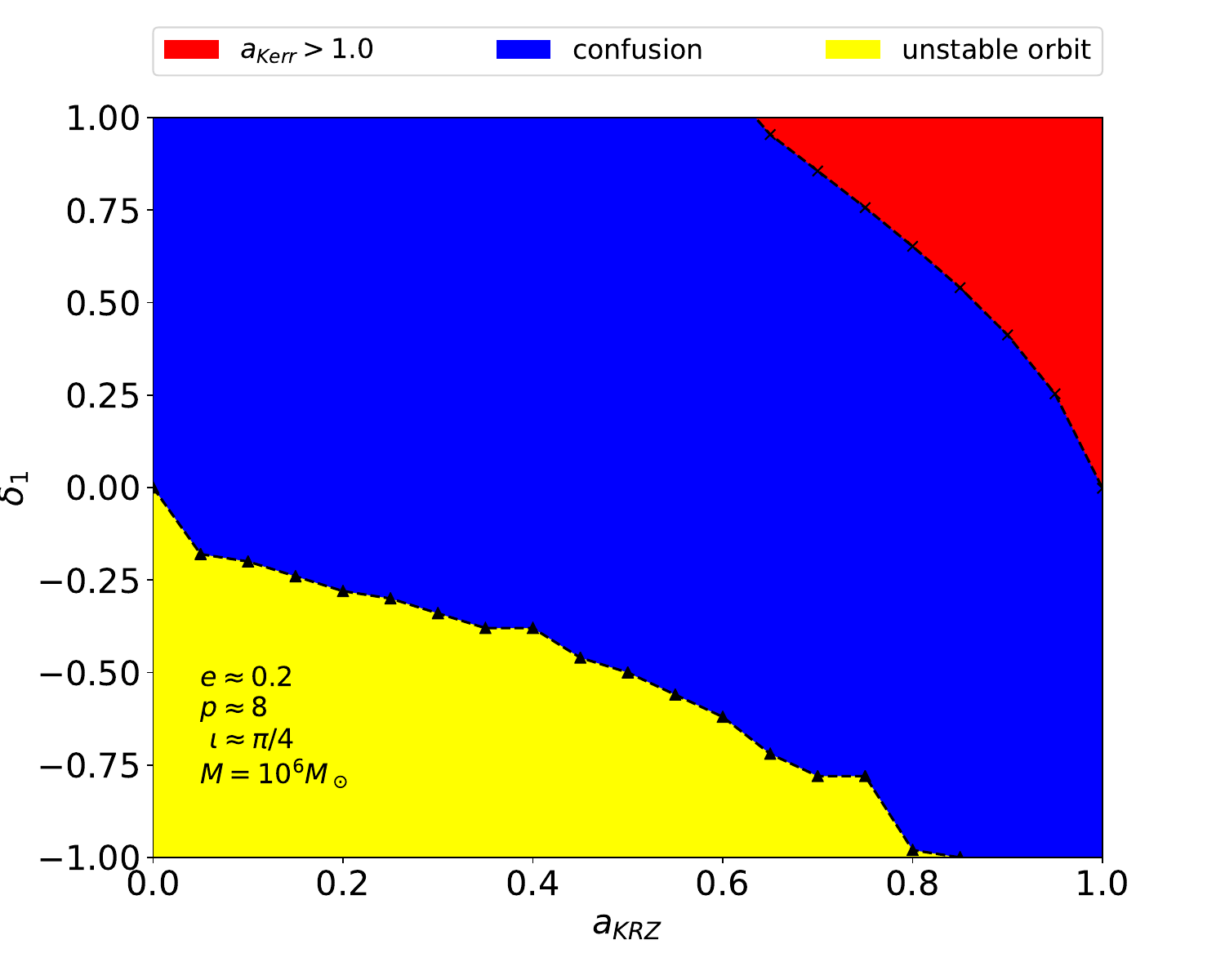}	
	\caption{The same as Fig. \ref{d2limit}, but for inclined orbit and waveform at each point is determined by setting initial $E,Lz$ same as corresponding Kerr orbit, see text for details.}
	\label{3dlimit}
\end{figure}

Similar to equatorial cases, requirements of $a_{\rm{Kerr}}<1$ and stable orbits set bound to the deformation parameters. However, since there's no carter-like constant in KRZ metric, we cannot control orbit parameters $(e,\, p,\, \iota)$ in non-Kerr cases in simple ways. Therefore, we try just to control the initial condition so that the orbit parameter are at least near the desired value, e.g. $(e,\, p,\, \iota) = (0.2,\, 8,\, \pi/4 )$. Technically, given $\delta_i$, $a_{\rm{KRZ}}$ and reference orbit parameters, we calculate the energy $E_{\rm{Kerr}}$ and angular momentum $L_{z -\rm Kerr}$ in Kerr spacetime with spin equal to $a_{\rm{KRZ}}$, then set the initial coordinate as $\theta=\pi/2$, $\phi=0$, t=0 and r at $\frac{p}{1-e}$, set $E,\, L_z$ equal to the Kerr value, which determines velocity in $t$ and $\phi$ direction, set $u^r=0$ and $u^\theta$ is determined by modulus of 4-velocity. The upper and lower bound of deformation parameters determined in this way is shown in Fig. \ref{3dlimit}. $M_{\rm{KRZ}}$ is $10^6$ solar mass and total time is $2\times 10^6$s.

\section{Conclusion and Discussion}
\label{p_fin}

LIGO's detections of GW signals have opened up the era of gravitational wave physics and astronomy. Future LISA task will be able to extend our sight over a broader spectrum in GW signals. One scientific goal of LISA task is to test Kerr metric or No-Hair Theorem in strong field region, but the ``confusion problem" is still a burden. In this paper, we consider the general parametrized metric of axisymmetric BHs, namely KRZ metric. With Kludge method for waveform generation, we investigate the confusion between EMRI waveforms with Kerr metric and KRZ non-Kerr metric. We mainly consider deformation parameters $\delta_1$ and $\delta_2$ which represent the deformation of Kerr metric components $g_{tt}$ and $g_{rr}$ respectively. For both equatorial and inclined orbits, we study the overlap between waveforms of same orbital frequencies.

The results show that the confusion exist in a large range of parameter space for both equatorial and inclined orbits, within the small and medium deviation ($\delta_i<1$) region. However, for high spin and deviation $\delta_i>0$, it is still possible to distinguish the back ground metric by physical restriction, $a_{\rm{Kerr}}<1$. In equatorial orbit cases, for a given orbital parameters $(e,p)$, the increase of $\delta_i$ is almost equivalent to proportional adding spin and mass of BH in Kerr metric, if just investigating the behavior of waveforms\comment{as far as waveforms are considered}. It means that in most cases, an EMRI waveform generated with KRZ metric can be mimicked by waveform templates with the Kerr black hole. This induces that one may not recognize the deviation from the No-Hair theorem. Therefore, when using EMRI signals by matched filtering method to test the Kerr spacetime, care must be taken to check the ``confusion problem''. 

But the ``confusion problem'' we have considered does not rule out the possibility of discerning BH metric with LISA detection. We have already demonstrated that if the spin of black hole is high enough, there are still a chance to distinguish the deviations from Kerr spacetime. Moreover, we did not consider the radiation reaction in the present paper. We expect that by a longer period of observation, where radiation reaction plays a more significant role, we could possibly observe the difference in orbit evolution by studying EMRI signal. For EMRIs with confusion problem from inclined orbits with same orbital frequencies, the semi-latus $p$ is different for Kerr and non-Kerr cases. Therefore the radiation flux of the GW is different and this can lead to different orbital evolution over longer time scale. Also, the confusion may be a result of approximation of waveform generation methods, which ignore higher order contributions to GW signals. In the future, we will extend our analysis to waveforms generated by more accurate methods and add radiation reaction over longer time scales. 

Another way out is multi-messenger measurements. It has been understood for long that X-ray emission, e.g. ironline, can be used to measure BH spin\cite{measureBHspin}, assuming Kerr hypothesis. With ironline, we are also able to put constraints on alternative metric. There has been some work constraining deformation in JP metric \cite{t1_Iron}, Johannsen metric \cite{t4_Iron}, KRZ metric \cite{cosimoKRZ} and etc. EMRI is extremely sensitive to small deviations, i.e. 0.1\% deviation of parameters could lead to significant change in overlap, but with the confusion problem, graphically, the confidence level will be an infinitely long tube with narrow openings. By combining ironline data and EMRI signal, we can possibly avoid confusion in the analysis. Naively thinking, while we can match the EMRI signal from non-Kerr BH of parameter $(\delta_i,a_{\rm KRZ},M_{\rm KRZ})$ with a Kerr signal $(0,a_{\rm Kerr}, M_{\rm Kerr})$, the parameter $(0,a_{\rm Kerr}, M_{\rm Kerr})$ might lies out of $3\sigma$ level in ironline fitting.

\bibliographystyle{unsrt}

\end{document}